\documentstyle[preprint,aps,epsfig]{revtex}
\begin{document}
\draft
\title{
{\begin{flushright}
{\footnotesize ZU-TH 24/01}\\
{\footnotesize YU-PP-I/E-KM-2-01}
\end{flushright}}
$m_u+m_d$ From Isovector Pseudoscalar Sum Rules}
\author{Kim Maltman\thanks{e-mail: kmaltman@physics.adelaide.edu.au;
permanent address: Department of Mathematics and Statistics, 
York University, 4700 Keele St., Toronto, Ontario, CANADA M3J 1P3}}
\address{CSSM, University of Adelaide, Australia 5005 and
Theory Group, TRIUMF, 4004 Wesbrook Mall, Vancouver, B.C.,
CANADA, V6T 2A3}
\author{Joachim Kambor\thanks{e-mail: kambor@physik.unizh.ch}}
\address{Institut f\"ur Theor. Physik, Univ. Z\"urich,
CH-8057 Z\"urich, Switzerland}
\maketitle
\begin{abstract}
We revisit the isovector pseudoscalar sum rule determination of $m_u+m_d$,
using families of finite energy sum rules
known to be very accurately satisfied in the isovector vector channel.  
The sum rule constraints are sufficiently strong 
to allow a determination of both $m_u+m_d$ and the
excited resonance decay constants.  The corresponding
Borel transformed sum rules are also very
well satisfied, providing a non-trivial consistency
check on the treatment of direct instanton contributions.  We obtain 
$[m_u+m_d](2\ {\rm GeV})=7.8\pm 1.1\ {\rm MeV}$ (in the
$\overline{MS}$ scheme), only marginally compatible
with the most recent sum rule determinations, but in good agreement with 
recent unquenched lattice extractions.
\end{abstract}

\section{Introduction}
Because of the Ward identity 
$\partial_\mu A_{ud}^\mu\, =\, \left( m_u+m_d\right) \bar{u}i\gamma_5 d$,
a study of the correlator
\begin{equation}
\Pi_{ud}(q^2) = i\, \int\, d^4x\, e^{iq\cdot x}\,
\langle 0\vert T\left( \partial_\mu A^\mu_{ud}(x) 
\partial_\nu {A^\nu_{ud}}^\dagger (0)\right) \vert 0\rangle
\label{pscorrelator}
\end{equation}
allows one, in principle, to determine $m_u+m_d$~\cite{v78}.
A number of such studies have been 
performed~\cite{v78,pssr,bpr,p98},
the most recent (Refs.~\cite{bpr} (BPR) and \cite{p98} (P98)) 
employing, respectively, 3- and 4-loop expressions for 
the dominant $D=0$ OPE contribution. 
We concentrate on the results of P98 (which updates BPR) in what follows.
P98 quotes, for $m_u+m_d${\begin{footnote}{All quark masses, 
here and in what follows, are in the $\overline{MS}$ scheme.}\end{footnote}}
\begin{equation}
[m_u+m_d](2\ {\rm GeV})=9.8\pm 1.9\ {\rm MeV}\ .
\label{p98mumd}
\end{equation}
Recent unquenched lattice simulations, in contrast, yield~\cite{CPPACS,QCDSF}
\begin{eqnarray}
&&[m_u+m_d](2\ {\rm GeV})=6.88^{+.28}_{-.44}\ {\rm MeV}\qquad {\rm (CP-PACS)}
\nonumber \\ 
&&[m_u+m_d](2\ {\rm GeV})=7.0\pm 0.4\ {\rm MeV}\qquad {\rm (QCDSF-UKQCD)}
\ ,
\label{latticemumd}\end{eqnarray}
where the errors do not reflect the uncertainty involved in using
perturbative versions of the renormalization constants. 
Because the consistency of the lattice and sum rule determinations
is not particularly good, we revisit the sum 
rule treatment of $\Pi_{ud}$.

In this paper, we study $\Pi_{ud}$ using Borel transformed {\it and} finite
energy sum rules (BSR's and FESR's).  The BSR's have the form~\cite{svz}
\begin{eqnarray}
&&M^6\, {\cal B}\left[ \Pi^{\prime\prime}_{ud}\right](M^2)\, =\,
\int_0^\infty\, ds\, e^{-s/M^2}\rho_{ud}(s)\label{borel} \\
&&\qquad\qquad\qquad\simeq\, \int_0^{s_0}\, ds\, e^{-s/M^2}\rho_{ud}(s)+
\int_{s_0}^\infty\, ds\, e^{-s/M^2}\rho^{OPE}_{ud}(s)\ ,
\label{bsr}
\end{eqnarray}
where $M$ is the Borel mass, $s_0$  the ``continuum
threshold'', and ${\cal B}\left[ \Pi^{\prime\prime}_{ud}\right](M^2)$ 
the Borel transform of $\Pi_{ud}^{\prime\prime}(Q^2)
\equiv d^2\Pi_{ud}(Q^2)/\left( dQ^2\right)^2$).
The FESR's have the form
\begin{equation}
{\frac{-1}{2\pi i}}
\oint_{\vert s\vert =s_0}\, ds\, w(s)\, \Pi_{ud}(s)\, = \,
\int_0^{s_0}\, ds\, w(s)\, \rho_{ud}(s)\ ,\label{basicfesr}
\end{equation}
where $s_0$ is arbitrary and 
$w(s)$ is any function analytic in the region of the contour.

The $\pi$ contribution to $\rho_{ud}$,
$\left[ \rho_{ud}(s)\right]_\pi = 2f_\pi^2 m_\pi^4 \delta \left(
s-m_\pi^2\right)$, with $f_\pi =92.4\ {\rm MeV}$, 
is very accurately known.  The decay
constants of the $\pi (1300)$ and $\pi (1800)$, needed
to describe the remaining contributions to $\rho_{ud}$
below $s\sim 4\ {\rm GeV}^2$, are not known, and
need to be determined as part of the sum rule analysis.

The LHS of Eq.~(\ref{borel}) can
be evaluated using the OPE, provided that $M$
is sufficiently large compared to the QCD scale.  The 
condition that $s_0$ be similarly large, though necessary, is not
sufficient for the OPE to be employed reliably on
the LHS of Eq.~(\ref{basicfesr}) since, except for
extremely large $s_0$, the OPE is expected to break down
near the timelike real axis~\cite{pqw}. 
For the isovector vector (IVV) channel, this breakdown can be
seen explicitly using the very precise spectral data available
from hadronic $\tau$ decay~\cite{ALEPHOPAL}: 
FESR's involving $w(s)=s^k$ with $k=0,1,2,3$
(which fail to suppress contributions
from the region of the circle $\vert s\vert =s_0$
near the timelike real axis) are rather
poorly satisfied at scales $2\ {\rm GeV}^2<s_0<m_\tau^2$\cite{kmfesr}.
The breakdown of the OPE, however, turns out to be
very closely localized to the vicinity of the
timelike axis: FESR's based on
weights having even a single zero at $s=s_0$
are very accurately satisfied over this whole range~\cite{kmfesr}.
Thus at the scales
$2\ {\rm GeV}^2<s_0<4\ {\rm GeV}^2$ of interest to us,
the supplementary constraint $w(s_0)=0$ must be imposed in order to 
obtain reliable FESR's.  We call such FESR's 
``pinch-weighted'', or pFESR's.

The OPE representation of $\Pi_{ud}(Q^2)$ is
known up to dimension $D=6$, with the dominant $D=0$ perturbative
contribution known to 4-loop order\cite{jm,cps}.  
Working with $\Pi^{\prime\prime}(Q^2)$, which allows
logarithms to be summed via the scale
choice $\mu^2 =Q^2$, one has~\cite{jm,cps}
\begin{eqnarray}
\left[ \Pi_{ud}^{\prime\prime}(Q^2)\right]_{D=0}\,& =&\,
{\frac{3}{8\pi^2}}{\frac{\left( \bar{m}_u+\bar{m}_d\right)^2}{Q^2}}
\left( 1+{\frac{11}{3}}\bar{a} + 14.1793 \bar{a}^2 + 77.3683 \bar{a}^3
\right)\label{psd0}\\
\left[ \Pi_{ud}^{\prime\prime}(Q^2)\right]_{D=4}\, &=&\,
{\frac{\left( \bar{m}_u+\bar{m}_d\right)^2}{Q^6}}
\Biggl( {\frac{1}{4}}\Omega_4
+{\frac{4}{9}}\bar{a}\Omega_3^{ss}
-\left[ 1+{\frac{26}{3}}\bar{a}\right] (m_u+m_d) <\bar{u}u >
\Bigg. \nonumber \\
&&\Bigg. \qquad -{\frac{3}{28\pi^2}}\bar{m}_s^4
\Biggr)\label{psd4} \\
\left[ \Pi_{ud}^{\prime\prime}(Q^2)\right]_{D=6}\, &=&\,
{\frac{\left( \bar{m}_u+\bar{m}_d\right)^2}{Q^8}}
\Biggl( -3\left[
\langle m_ug\bar{d}\sigma\cdot G d+m_dg\bar{u}\sigma\cdot G u\rangle
\right]\nonumber \Bigg. \\
\Bigg. &&\qquad -{\frac{32}{9}}\pi^2 a\rho_{VSA}
\left[ \langle \bar{u} u\rangle^2
+\langle \bar{d}d\rangle^2 - 9\langle \bar{u}u\rangle
\langle \bar{d}d\rangle\right]\Biggr)\ ,
\label{psd6}
\end{eqnarray}
where $\bar{a}\equiv a(Q^2)=\alpha_s(Q^2)/\pi$,
$\bar{m}_k\equiv m_k(Q^2)$, with $\alpha_s(Q^2)$ and 
$m(Q^2)$ the running coupling 
and running mass at scale $\mu^2=Q^2$ in the $\overline{MS}$ scheme,
$\Omega_4$ and $\Omega_3^{ss}$ are the 
RG invariant modifications
of $\langle a G^2\rangle$ and $\langle m_s\bar{s}s\rangle$ defined
in Ref.~\cite{jm}, and $\rho_{VSA}$ in Eq.~(\ref{psd6})
describes the deviation of the four-quark condensates
from their vacuum saturation values.  We have dropped
$D=2$ contributions, which are suppressed by 
two additional powers of $m_{u,d}$, and additional
$D=4$ contributions proportional to $[m_u+m_d]^2 m_{u,d}^4$.
The Borel transforms
of the above expressions are well known, and may be found in
Refs.~\cite{jm,cps}.

In scalar and pseudoscalar channels, 
direct instanton contributions are potentially important,
but are not incorporated
in the OPE representation of $\Pi_{ud}$~\cite{novikov81,shuryak8283}. 
We estimate their size using the instanton liquid model~\cite{ilm}.
ILM contributions to
the theoretical side of the $\Pi_{ud}$ BSR are given by
\begin{equation}
{\frac{3\rho_I^2\left( m_u+m_d\right)^2 M^6}{8\pi^2}}\left[
K_0(\rho_I^2 M^2/2)+K_1(\rho_I^2 M^2/2)\right] ,
\label{spsbsrinst}
\end{equation}
where $\rho_I\simeq (1/0.6\ {\rm GeV})$ is the average
instanton size and $K_i$ are the MacDonald functions.  ILM contributions 
play only a small (few percent) role in the BSR analysis 
at scales $M^2>2\ {\rm GeV}^2$, but are important
for FESR analyses.
For polynomial weights, ILM FESR contributions
follow from~\cite{elias98}
\begin{equation}
{\frac{-1}{2\pi i}}\, \oint_{\vert s\vert =s_0} ds\, s^k
\left[ \Pi_{ud}(s)\right]_{ILM} =
{\frac{-3[m_u+m_d]^2}{4\pi}}\, \int_0^{s_0} ds\, s^{k+1}J_1\left(
\rho_I\sqrt{s}\right) Y_1\left(\rho_I\sqrt{s}\right)\ .
\label{fesrinstanton}
\end{equation}

We employ the following values for OPE/ILM input:
$\rho_I=1/(0.6\ {\rm GeV})$~\cite{shuryak8283,ilm};
$\alpha_s(m_\tau^2)=0.334\pm .022$~\cite{ALEPHOPAL};
$\langle \alpha_s G^2\rangle =(0.07 \pm 0.01)\ {\rm GeV}^4$~\cite{narisonaGG};
$\left( m_u+m_d\right)\langle \bar{u}u\rangle =-f_\pi^2 m_\pi^2$;
$0.7< \langle \bar{s}s\rangle /\langle \bar{u} u\rangle 
\equiv r_c<1$~\cite{jm,cps};
$\langle g\bar{q}\sigma Fq\rangle
=\left( 0.8\pm 0.2\ {\rm GeV}^2\right)\langle \bar{q} q\rangle$\cite{op88};
and $\rho_{VSA}=0\rightarrow 10$.  The $D=0$ and $4$ OPE 
contributions are evaluated via the 
contour-improvement prescription~\cite{cipt1}, using
the analytic solutions for $\alpha_s(Q^2)$ and $m(Q^2)$ obtained from
the 4-loop-truncated versions of the $\beta$~\cite{beta4} 
and $\gamma$~\cite{gamma4} functions.

\section{Potential problems with the existing sum rule treatment
and the updated pFESR analysis}
The BPR and P98 analyses employ FESR's with $w(s)=1,s$.  The
global normalization of the resonance contributions is fixed by
assuming resonance dominance of $\rho_{ud}$ at $3\pi$ threshold and
normalizing the tails of the resonance contributions 
to the known ChPT threshold expression.  
Direct instanton contributions are neglected.
The relative strengths of the two resonance contributions
are constrained by optimizing a ``duality'' match between OPE and
spectral ansatz versions of the ratio of the $w(s)=1$-
and $s$-weighted FESR's{\begin{footnote}{The duality
matching is equivalent to imposing the local duality (OPE)
version of the spectral function, up to an overall multiplicative
constant, over the whole of the matching window.}\end{footnote}}.
$m_u+m_d$ is then extracted from the optimized duality matching
region of the $w(s)=1$-weighted FESR.  The result of Eq.~(\ref{p98mumd})
corresponds to $s_0\sim 2\ {\rm GeV}^2$.  The determination
is, in fact, not stable with respect
to $s_0$, falling roughly linearly from $9.8\ {\rm MeV}$ 
at $s_0\simeq 2\ {\rm GeV}^2$ to $7.5\ {\rm MeV}$ 
at $s_0\simeq 4\ {\rm GeV}^2$ (see Fig.~2 of P98).

Two potential problems with this analysis are (1) the
use of non-pinched-weighted FESR's at scales for which they
are poorly satisfied in the IVV channel, and (2) neglect of 
direct instanton contributions.
In addition, the P98 result,
$[m_u+m_d](1\ {\rm GeV})/[m_u+m_d](2\ {\rm GeV})=1.31$ 
corresponds (using 4-loop running) to 
$\alpha_s(m_\tau^2)=0.307$, significantly
lower than the recent ALEPH determination.  Since this will produce
an overestimate of $m_u+m_d$, an update of OPE input is also in
order.

\begin{figure}
\centering{\
\psfig{figure=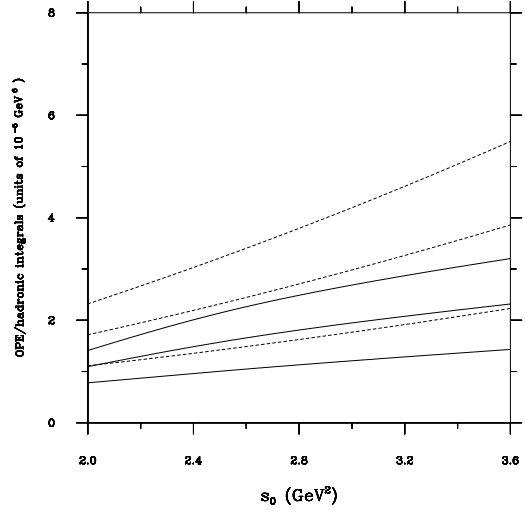,height=7.5cm}}
\vskip 0.15in
\caption{The $w_N$ OPE/spectral integral match 
corresponding to central values of all OPE input, the quoted P98 value  
$[m_u+m_d](1\ {\rm GeV})=12.8\ {\rm MeV}$ and the P98 spectral ansatz.
The solid (dashed) lines represent the spectral (OPE) integrals.  
The lower, middle and upper lines for each case
correspond to $A=0,2$ and $4$, respectively.}
\label{bprNps}
\end{figure}

Concerning the first problem, one could, of course, be lucky:  
the scale at which the OPE
can be safely used right down to the timelike 
axis might turn out to be lower in the isovector pseudoscalar than in the 
IVV channel.  If so, however, pFESR's employing
the same spectral ansatz and same value of $m_u+m_d$ should
also be well satisfied at the scales used in P98.  
We test this possibility using pFESR's based on 
$w_N(y,A)=(1-y)(1+Ay)$ and $w_D(y,A)=(1-y)^2(1+Ay)$
(where $y=s/s_0$ and $A$ is a free parameter), which are known
to be well satisfied in the IVV channel.
The resulting OPE/spectral integral match (corresponding
to the above values for the OPE input and, as in P98, neglect of
ILM contributions) is shown for the $w_N$ case in Fig.~\ref{bprNps},
and is obviously quite poor.
(The quality of the $w_D$ match is even worse.)

\begin{figure}
\centering{\
\psfig{figure=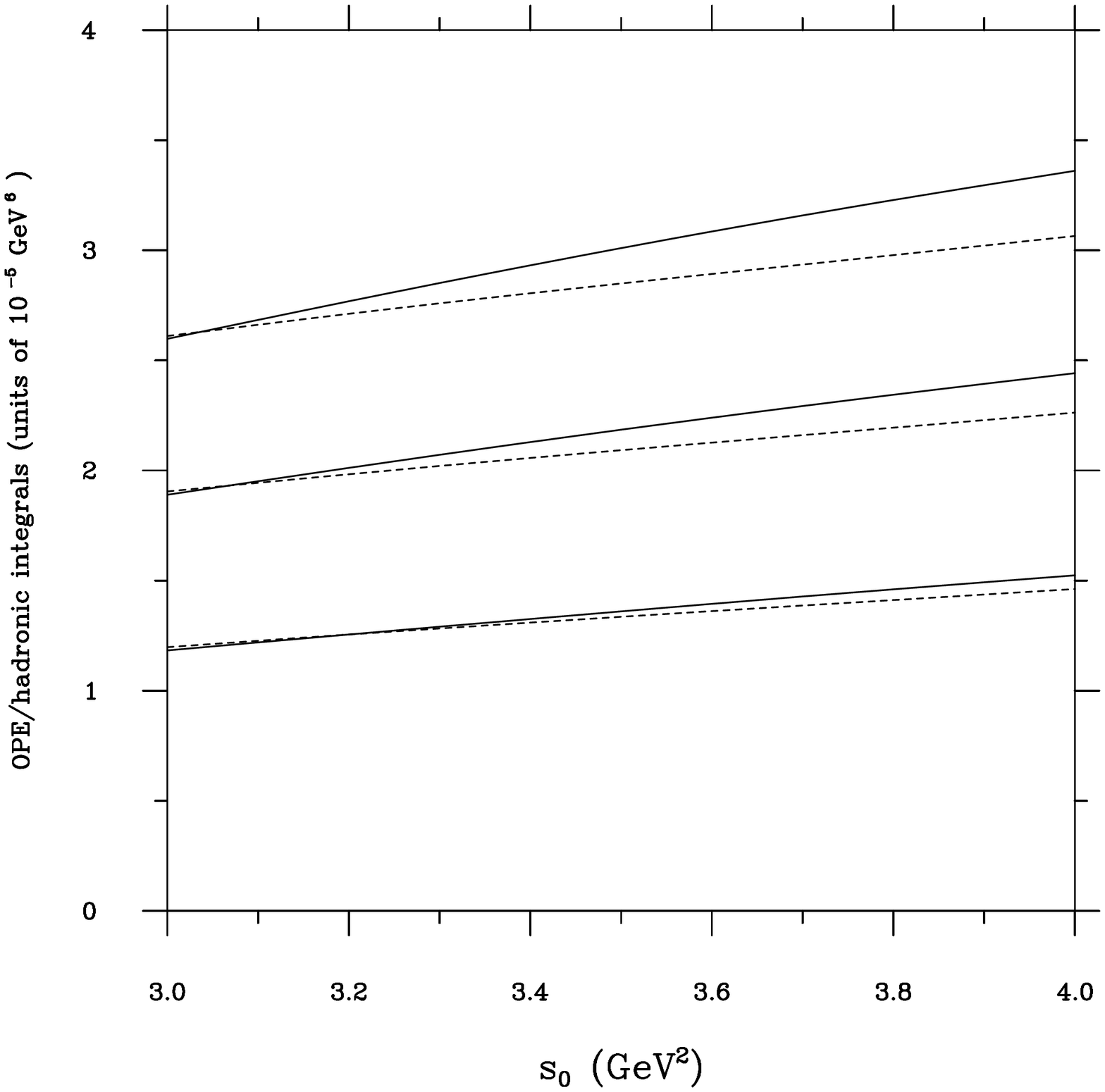,height=7.5cm}}
\vskip 0.15in
\caption{The $w_N$ OPE+ILM/spectral integral match corresponding to
central values of all OPE input, the ILM estimate of direct
instanton contributions, and use of the P98 spectral ansatz.
The OPE+ILM curves employ the optimized value, 
$[m_u+m_d](2\ {\rm GeV})=6.8\ {\rm MeV}$,
obtained in a combined $w_N$, $w_D$ pFESR fit.
The conventions for identifying spectral and OPE+ILM integrals,
and the $A=0,2$ and $4$ cases, are as for Fig.\ref{bprNps} above.}
\label{bprilmfesrN}
\end{figure}

\begin{figure}
\centering{\
\psfig{figure=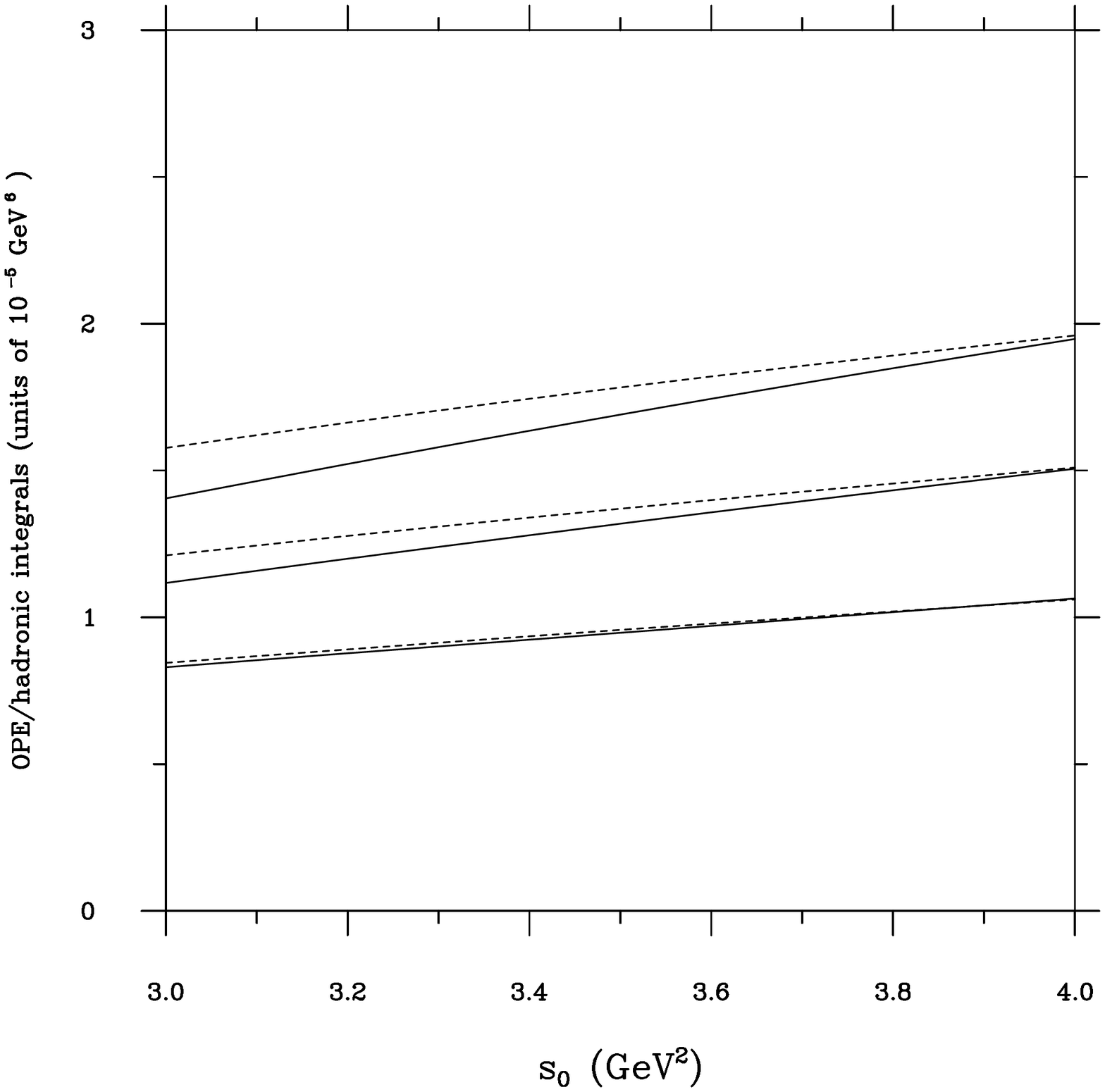,height=7.5cm}}
\vskip 0.15in
\caption{The OPE+ILM/spectral integral match, as in Fig.\ref{bprilmfesrN},
except for the $w_D$ rather than $w_N$ weight family.}
\label{bprilmfesrD}
\end{figure}

We have re-analyzed $\Pi_{ud}$, using the P98 spectral ansatz as
input, but fixing $m_u+m_d$ via a combined 
$w_N$, $w_D$ pFESR analysis.  If we do not include
ILM contributions, the optimized OPE/spectral integral
match remains poor.  Including ILM contributions 
produces a reasonable optimized match.  The corresponding 
value of $m_u+m_d$ is
\begin{equation}
[m_u+m_d](2\ {\rm GeV})=6.8\ {\rm MeV}\ .
\label{p98reanalysisilm}
\end{equation}
The quality of this match is shown 
in Fig.~\ref{bprilmfesrN} for the $w_N$ family 
and in Fig.~\ref{bprilmfesrD} for the $w_D$ family.
The result of Eq.~(\ref{p98reanalysisilm}) is compatible with the P98 
results corresponding to $s_0\simeq 4\ {\rm GeV}^2$ 
but significantly smaller than that corresponding to  $s_0=2\ {\rm GeV}^2$.
Since, in spite of optimization, the match for $w_N$ is best where that for
$w_D$ is worst, and vice versa, it appears that some modification of the P98
spectral ansatz is also required.

\begin{figure}
\centering{\
\psfig{figure=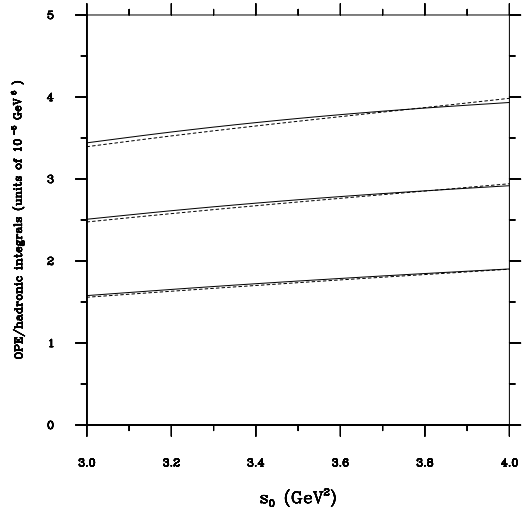,height=7.5cm}}
\vskip 0.15in
\caption{The optimized OPE+ILM/spectral integral match
for the $w_N$ pFESR family, with $m_u+m_d$, $f_1$ and $f_2$
given by the central values of Eqs.~(\ref{udilmmass}),
(\ref{udilmf1}) and (\ref{udilmf2}).  The labelling of the
hadronic integrals, OPE integrals and the 
$A=0,2$ and $4$ cases, is as for Fig.\ref{bprNps} above.}
\label{figudNILM}
\end{figure}

\begin{figure}
\centering{\
\psfig{figure=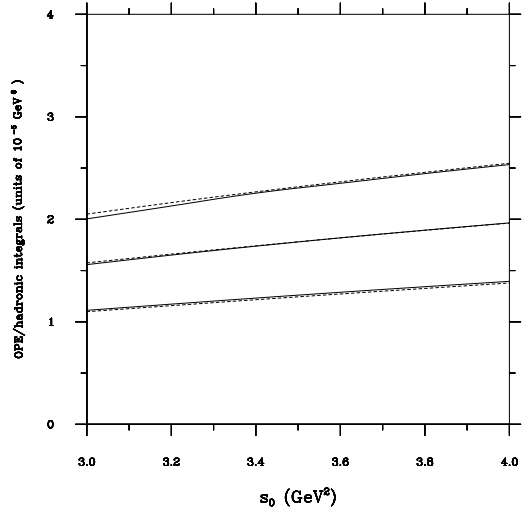,height=7.5cm}}
\vskip 0.15in
\caption{The optimized OPE+ILM/spectral integral match,
as in Fig.~\ref{figudNILM}, except for the $w_D$ rather
than $w_N$ weight family.}
\label{figudDILM}
\end{figure}

In this work, we aim to determine simultaneously the excited resonance
decay constants, $f_{\pi (1300)}\equiv f_1$
and $f_{\pi (1800)}\equiv f_2$, which characterize the
modifications of the spectral ansatz, and $m_u+m_d$.
To this end, we perform a combined
$w_N$ and $w_D$ pFESR analysis{\begin{footnote}{The same
treatment of the IVV channel results in a
determination of $f_\rho$ accurate to within the
experimental error~\cite{kma0}.}\end{footnote}}.  Our spectral ansatz is
\begin{equation}
\rho_{ud} (s)\, =\, 2f_\pi^2m_\pi^4\delta \left( s-m_\pi^2\right)
+2f_1^2m_{\pi (1300)}^4 B_1(s) 
+2f_2^2m_{\pi (1800)}^4 B_2(s)\ ,
\label{rhoform}
\end{equation}
where $B_{1,2}(s)$ are standard Breit-Wigner forms
for the $\pi (1300)$ and $\pi (1800)$.
We employ
PDG2000~\cite{pdg00} values for the masses and widths.
This ansatz can be used sensibly only up to
$s_0\simeq \left( m_{\pi (1800)}+\Gamma_{\pi (1800)}\right)^2
\simeq 4\ {\rm GeV}^2$.
To maintain good convergence of the OPE
we also require $s_0\geq 3\ {\rm GeV}^2$.
For $3\ {\rm GeV}^2\leq s_0 \leq 4\ {\rm GeV}^2$,
the integrated $D=0$ OPE series converges well for all $A\geq 0$.
Larger $A$ produces
larger relative contributions
from the resonance region, and hence aids in the
extraction of $f_1$ and $f_2$.
The results of this analysis are
\begin{eqnarray}
&&[m_u+m_d](2\ {\rm GeV})\, =\, 7.8\pm 0.8_\Gamma 
\pm 0.5_{theory}\pm 0.4_{method}\ 
{\rm MeV}\label{udilmmass} \\
&&f_1\, =\, 2.20\pm 0.39_{\Gamma}\pm 0.18_{theory}\pm 0.18_{method}
\ {\rm MeV}\label{udilmf1} \\
&&0< f_2< 0.37\ {\rm MeV}\ .
\label{udilmf2}
\end{eqnarray}
The errors labelled ``$\Gamma$'' result from varying the input
resonance parameters within the PDG2000 errors, and are due
essentially entirely to the (large) uncertainty on the
$\pi (1300)$ width.  Those labelled ``theory'' reflect 
uncertainties in the OPE input and our
estimate of the error associated with truncating
the $D=0$ series at $O(a^3)$.  Those labelled ``method''
are obtained by studying the impact of employing different
analysis windows in $s_0$ and $A$, and performing
separate $w_N$ and $w_D$ analyses.  Further details of the 
analysis, and a breakdown of the separate error contributions
will be given elsewhere~\cite{kmjkbigpssr}.
The OPE+ILM/spectral integral match corresponding to these
results, shown in Fig.~\ref{figudNILM} for the
$w_N$ family and Fig.~\ref{figudDILM} for the $w_D$ family,
is obviously excellent.

\begin{figure}
\centering{\
\psfig{figure=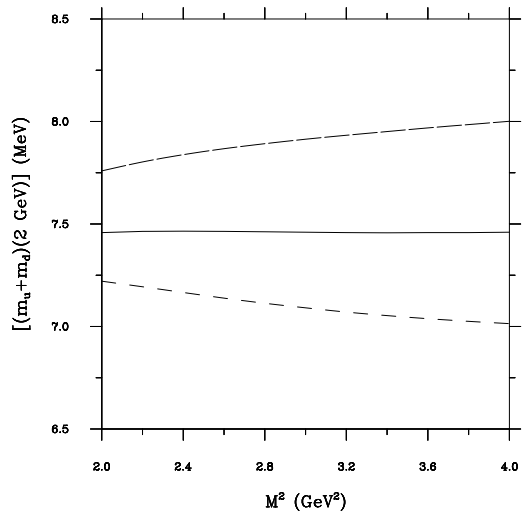,height=7.5cm}}
\vskip 0.15in
\caption{$[m_u+m_d](2\ {\rm GeV})$, as a function
of $M^2$ for the BSR analysis described in the text.  The solid
line corresponds to $s_0=3.7\ {\rm GeV}^2$, which produces
optimal stability for $m_u+m_d$ with respect to $M^2$ in the window 
$2\ {\rm GeV}^2\leq M^2\leq 3\ {\rm GeV}^2$.
The lower (short) dashed line corresponds
to $s_0=4.2\ {\rm GeV}^2$ and the upper
(long) dashed line to $s_0=3.2\ {\rm GeV}^2$.  Note the compressed
vertical scale.}
\label{udborel}
\end{figure}

As noted above, the ILM contributions play a non-negligible
role in the pFESR analysis.  In fact, if one removes 
ILM contributions, an equally good 
OPE/spectral integral match is obtained, but now corresponding to
$[m_u+m_d](2\ {\rm GeV})\, =\, 9.9\pm 1.2_\Gamma \pm
1.0_{theory}\pm 0.5_{method}\ {\rm MeV}$, 
$f_1=2.41\pm 0.50_{\Gamma}\pm 0.21_{theory}\pm 0.27_{method}\ {\rm MeV}$ and
$f_2=1.36\pm 0.16_{\Gamma}\pm 0.09_{theory}\pm 0.11_{method}\ {\rm MeV}$.  
The pFESR analysis alone thus
provides no evidence either for or against
including ILM contributions.  Fortunately, the requirement of consistency 
between BSR and pFESR analyses places non-trivial constraints
on the ILM representation.  This works as follows.   The pFESR analysis
provides a determination of $m_u+m_d$ and $f_{1,2}$ which is sensitive
to whether or not ILM contributions are included.  The output
$f_{1,2}$, together with $f_\pi$, determine
the low-$s$ part of $\rho_{ud}$, and hence can be
used as input to a BSR analysis.  The high-$s$
part is, as usual, approximated by the continuum
ansatz, with the continuum threshold, $s_0$, determined by 
optimizing stability of the BSR output (in this case $m_u+m_d$)
with respect to $M$.  The BSR and pFESR output values for $m_u+m_d$
should be compatible 
if the ILM representation is reasonable.  Errors 
associated with uncertainties in input OPE and resonance parameter
values are common to the pFESR and BSR analyses and strongly
correlated.  Additional errors are present for the BSR
analysis as a result of the crudeness of the continuum approximation
and the uncertainties in the criterion for fixing $s_0$.
We assign a $20\%$ error to continuum spectral contributions,
and allow $s_0$ to vary by
$\pm 0.5\ {\rm GeV}^2$ about the optimal stability value.
We work in a window of Borel masses $2\ {\rm GeV}^2\leq M^2
\leq 3\ {\rm GeV}^2$ for which continuum contributions are
$<50\%$ of the dominant $D=0$ OPE contribution and 
convergence of the Borel transformed $D=0$ OPE series 
is still good.  For the case that the ILM contributions are included in the
pFESR analysis we obtain, quoting only the {\it additional}
errors present for the BSR analysis,
\begin{equation}
[m_u+m_d](2\ {\rm GeV})=7.5\pm 0.9\ {\rm MeV}\ ,
\label{bsrilmmumd}
\end{equation}
to be compared to the {\it central} value given 
in Eq.~(\ref{udilmmass}).  The agreement is excellent.
The stability of the BSR analysis, shown in Fig.~\ref{udborel},
is also extremely good.  In contrast, if ILM contributions
are omitted from the pFESR analysis, the BSR result
becomes $[m_u+m_d](2\ {\rm GeV})=8.8\pm 0.6\ {\rm MeV}$,
incompatible with the pFESR determination.  If one performs a pFESR
optimization of $f_{1,2}$ separately for each $m_u+m_d$,
one finds that, with no ILM contributions,
the pFESR and BSR values remain
inconsistent, within the additional BSR errors, unless the
pFESR input, $[m_u+m_d](2\ {\rm GeV})$, is $<8.1\ {\rm MeV}$.
The corresponding optimized value for $f_1$ for this
marginal case turns out to be consistent within errors with
that quoted in Eq.~(\ref{udilmf1}), though the OPE/spectral
integral match is significantly worse than that obtained for
the optimized fit, including ILM contributions.  The low
value for $m_u+m_d$ thus appears to be an unavoidable
feature of the combined analysis of $\Pi_{ud}$.

\section{Summary and Discussion}
Combining the $0.3\ {\rm MeV}$ difference of pFESR and BSR
central values in quadrature with
all other sources of error,
we obtain, for our final result,
\begin{equation}
[m_u+m_d](2\ {\rm MeV})=7.8\pm 1.1\ {\rm MeV}\ .
\label{finalmumd}
\end{equation}
This is compatible, within errors, with the unquenched
lattice determinations of Refs.~\cite{CPPACS,QCDSF}, and
with the result obtained by combining the ChPT determination
$R\equiv 2m_s/[m_u+m_d]=24.4\pm 1.5$~\cite{leutwylermq}
with recent determinations of $m_s$ using
hadronic $\tau$ data~\cite{mstau}.
A recent summary~\cite{rgkm00} 
gives $83\ {\rm MeV}<m_s(2\ {\rm GeV})<130\ {\rm MeV}$,
which corresponds to $6.8<[m_u+m_d](2\ {\rm GeV})<10.7\ {\rm MeV}$.
Our analysis, in fact, favors values of $m_s$ in the lower
part of this range, in good agreement with recent
unquenched lattice results for $m_s$.


\acknowledgements
KM would like to acknowledge the ongoing support of the Natural Sciences and
Engineering Research Council of Canada, and also the hospitality of the
Theory Group and TRIUMF and the Special Research Centre for the 
Subatomic Structure of Matter at the University of Adelaide.  
JK would like to acknowledge partial support from Schweizerischer 
Nationalfonds.
\vskip .5in

\noindent

\vskip .5in

\vskip .5in

\vskip .5in

\vskip .5in\noindent

\end{document}